\documentclass[reprint,aps,prb,superscriptaddress,showkeys]{revtex4-2}
\usepackage[final]{graphicx}
\usepackage[utf8]{inputenc}
\usepackage[hidelinks]{hyperref}
\usepackage[normalem]{ulem}
\usepackage{bm,physics,amsmath,amssymb,xcolor,enumitem,tabularray,float,booktabs,placeins,siunitx}
\usepackage{changepage} 
\usepackage{makecell} 

\usepackage[export]{adjustbox}

\graphicspath{{./figures/}}

\begin{document}

\title{The R-index: A universal metric for evaluating OAM content\\ and mode purity in optical fields}

\author{Monika Bahl}
\email{monika.bahl@univie.ac.at}
\affiliation{Optical Metrology Group, Faculty of Physics, University of Vienna, Boltzmanngasse 5, 1090 Vienna, Austria}
\author{Georgios M. Koutentakis}
\email{georgios.koutentakis@ist.ac.at}
\affiliation{Institute of Science and Technology Austria (ISTA), Am Campus 1, 3400 Klosterneuburg, Austria}
\author{Mikhail Maslov}
\affiliation{Institute of Science and Technology Austria (ISTA), Am Campus 1, 3400 Klosterneuburg, Austria}
\author{Tom Jungnickel}
\affiliation{Optical Metrology Group, Faculty of Physics, University of Vienna, Boltzmanngasse 5, 1090 Vienna, Austria}
\author{Timo Ga\ss en}
\affiliation{Optical Metrology Group, Faculty of Physics, University of Vienna, Boltzmanngasse 5, 1090 Vienna, Austria}
\author{Mikhail Lemeshko}
\affiliation{Institute of Science and Technology Austria (ISTA), Am Campus 1, 3400 Klosterneuburg, Austria}
\author{Oliver H. Heckl}
\email{oliver.heckl@univie.ac.at}
\affiliation{Optical Metrology Group, Faculty of Physics, University of Vienna, Boltzmanngasse 5, 1090 Vienna, Austria}

\begin{abstract}

    Despite its pivotal role in optical manipulation, high‑capacity communications, and quantum information, a general measure of orbital angular momentum (OAM) in structured light remains elusive. In optical fields, where multiple vortices coexist, the local nature of vortex OAM and the absence of a common rotation axis make the total OAM of the field difficult to quantify. Here, we introduce the R‑index---a metric that captures the intrinsic OAM content of any structured optical field, from pure Laguerre-Gaussian modes to arbitrary multi‑vortex superpositions. Not only does this metric quantify the total OAM, it also assesses field purity, providing insight into the fidelity and robustness of the OAM generation. By unifying OAM characterization into a single figure of merit, the R‑index enables direct comparison across diverse beam profiles and facilitates the identification of optimal configurations for both foundational studies and applied technologies.
\end{abstract}


\maketitle

\section{Introduction}
\label{sec_introduction}
{Since the seminal work of Allen \emph{et al.} \cite{Allen_1992}, which demonstrated that electromagnetic waves can carry well‑defined orbital angular momentum (OAM), this discrete, and in principle unbounded, property of light has evolved from a theoretical concept into a practical resource. The ladder of OAM eigenstates offers high‑dimensional encoding capacity for classical and quantum communication \cite{Willner2015,Yao:11,yang2023high}, facilitates the transfer of rotational torque in optical micromanipulation \cite{He1995,Volpe2023}, and underpins emerging OAM-enhanced spectroscopic techniques~\cite{Maslov_2024}. Recent advances in wavefront-shaping optics, spiral phase plates, computer‑generated holograms, and spatial light modulators, now permit the routine generation and analysis of OAM beams \cite{singularities,senthilkumaran2024singularities,AwasthiKang2022,RosalesGuzman2024}, thereby broadening their applicability across photonics and related disciplines \cite{ANDREWS2004133,Babiker2002,Andrews2012,ForbesAndrews2021}}, also including quantum information processing \cite{Zhou2016OAM}, particle manipulation \cite{Molina-Terriza2007}, and light-matter interaction \cite{PhysRevLett.89.143601,PhysRevLett.122.103201,PhysRevA.86.023816,Schmiegelow2016,PhysRevLett.122.103201}.

The total angular momentum of an optical beam comprises an \textit{intrinsic} component, determined solely by the internal field distribution, and an \textit{extrinsic} component that depends on the displacement of the beam axis relative to the observation axis. Within the intrinsic part, one distinguishes spin angular momentum (SAM),related to the polarization of light, and OAM, which stems predominantly from the \textit{optical vortices} \cite{Allen_1992}. 
In a coherent light wave, represented by a complex scalar field propagating in free space, optical vortices arise at points of zero intensity, called nodes, where the phase is undefined. Around these singularities, the  phase gradient forms a circulating pattern, characterized by an azimuthal phase dependence of the form $\exp(il\phi)$, where $l$ is the topological charge and $\phi$ is the azimuthal angle. Such a phase structure in an optical vortex corresponds to a quantized angular momentum of magnitude $l \hbar$ per photon.
As clarified in Ref.~\cite{BliokhNori2015}, the helical phase factor $\exp(i l \phi)$ is the principal descriptor of the intrinsic OAM. It gives rise to a precession of the Poynting vector and the associated azimuthal orbital flow density (OFD). In contrast, SAM arises from the  rotation of electric field vectors and is quantified by the spin flow density (SFD) \cite{Bekshaev2011}. Together, OFD and SFD exhaust the intrinsic angular momentum content of an optical beam. These complementary flow densities offer experimentally accessible measures of the respective angular momentum contributions.

For an ideal paraxial beam containing a single vortex, all photons share the same helical phase factor, and the OAM can be unambiguously defined with respect to the common propagation axis. But, in practice, many beams are spatially structured optical fields (SSOFs) that contain multiple vortices. In such cases, three considerations preclude a straightforward  definition of the total orbital angular momentum:
\begin{enumerate}
\item {\it Local character of a vortex OAM.} Each vortex is associated with a unique phase singularity ,whose topological charge may differ from charges of other vortices. The associated OAM remains local to the singularity and does not extend globally on the beam.
\item {\it Vortex interference.} Superposition of individual vortex wavefronts generates intricate interference patterns and secondary phase singularities, such that a simple sum of single‑vortex OAM does not describe the composite field.

\item {\it Absence of a common rotation axis.} Whereas a single vortex beam, such as a Laguerre-Gaussian mode, has a single optical axis, the vortex array in a generic SSOF is described by several, in general different, rotational axes. Subsequently, the overall field lacks global rotational symmetry.
\end{enumerate}
Collectively, these factors imply that the total OAM content of an elaborately structured field cannot be deduced by enumerating vortices or by projecting onto a standard mode basis. Accessing the fitness of the field for applications, for instance, OAM‑enhanced spectroscopy \cite{Maslov_2024} requires a quantitative measure of the rotational energy flux in terms of a more rigorous metric.

In this paper, we introduce a unified measure, the R-index, based on the relative amplitude of the solenoidal component of the OFD, which provides an estimate of the photon fraction available for OAM‑mediated interactions. Our calculations demonstrate its efficacy in identifying the OAM content across a broad range of beam types, including Laguerre-Gaussian (LG), Gaussians, superpositions of Hermite-Gaussian (HG) beams and vortex-bearing optical lattices making it a universal tool for structured light analysis. The ability of the R-index to reduce the OAM content characterization to a single figure-of-merit makes it an attractive tool for selecting the appropriate structure of the electric field for OAM-enabled fundamental studies or technological applications. Moreover, we demonstrate its ability to act as a quality factor by assessing the mode purity of LG modes generated by spiral phase plates. This application is particularly important for mid-infrared applications where high-fidelity optical vortex generating apparatuses, such as spatial light modulators are unavailable.

Our paper is structured as follows. In Sec.~\ref{sec:OFD_characteristics} we review how the orbital flow density of a light field can be extracted from its phase gradients. In Sec.~\ref{HHD} we introduce our main tool the Helmholtz-Hodge decomposition, via which we can separate the solenoidal, OAM bearing, phase gradients from the irrotational part.
In Sec.~\ref{sec:R-index} we define the main quantity discussed in our paper, the R-index. Section~\ref{sec:numerics} summarizes the numerical techniques employed for the evaluation of the R-index. The focal parts of our work are Sec.~\ref{sec:results} describing our numerical findings for various field profiles and Sec.~\ref{sec:analytics} providing the analytical evaluation of the R-index for Laguerre-Gaussian beams. Sec.~\ref{sec:conclusion} summarizes our findings and provides future perspectives. Finally, Appendix~\ref{app:HGSSOFs} provides extended data tables for various superpositions of Hermite-Gaussian modes.

\section{Orbital flow characteristics}
\label{sec:OFD_characteristics}

Our approach consists of tracing the internal energy flows of the field, also known as optical currents, one of its intrinsic features that reveals details of its structure and dynamics \cite{Berry_2009}. These paths can be represented as curves that are everywhere tangent to the Poynting vector, a quantity that links to the dynamical attributes of optical fields. We consider a field $\psi({\bm r}) = |E({\bm r})| e^{i \varphi({\bm r})}$ and ${\bm E}({\bm r}) = \hat{\bm e}_p \psi({\bm r})$ and $\hat{\bm e}_p$ is the polarization of light, assumed here to be homogeneous so we can treat the electric field as scalar.

The optical current associated to the field $\psi({\bm r})$ is ${\bm j}({\bm r})$ \cite{Berry_2009}, it equals the time-averaged energy flow and is given by
\begin{equation}
    {\bm j}({\bm r}) = I({\bm r}) \boldsymbol\nabla\varphi({\bm r}),
    \label{current}
\end{equation}
where $I({\bm r}) = \frac{c \epsilon_0}{2} \psi({\bm r}) \psi^*({\bm r})$ is the intensity and $\psi^*({\bm r})$ is the conjugate of $\psi({\bm r})$, $\boldsymbol{\nabla}$ represents the gradient operator, and $\varphi({\bm r}) \equiv \text{arg}(\psi({\bm r}))$ is the phase of the optical field. 
The optical current is closely associated with the Poynting vector $\bm{P}({\bm r})$, indicating the flow of energy and is directly proportional to the canonical energy flow density. The Poynting vector reads
\begin{equation}
    \bm{P}(\bm{r}) = \frac{\Im[\psi^* \boldsymbol{\nabla} \psi]}{2 \mu_0 \omega}  = \frac{|\psi|^2 \boldsymbol{\nabla}\varphi({\bm r})}{2 \mu_0 \omega}  = \frac{1}{k} I({\bm r}) \boldsymbol{\nabla}\varphi({\bm r})\,,
    \label{poynting_vector}
\end{equation}
The right hand side of the above directly yields the relation ${\bm P}({\bm r}) = {\bm j}({\bm r})/k$.
This expression also shows the association of both ${\bm j}({\bm r})$ and ${\bm P}({\bm r})$ with the the probability current in quantum mechanics ${\rm Im}[\psi^*({\bm r}) \boldsymbol{\nabla} \psi({\bm r})]$ and which is sometimes used synonymously with the Poynting vector in electromagnetic fields.
$\bm{P}(\bm{r})$ is essential for calculating the OAM of any optical field and aids in understanding how small particles interact with the field \cite{McDonald2015OrbitalAS}.
This momentum density has a transverse azimuthal component produced by the helical phase of the optical vortex and is a signature of the vortex modes. It is canonical and proportional to the local gradient of the phase of the field, that is, to the local wave vector. Therefore, the canonical orbital energy flow is independent of the polarization in uniformly polarized fields and can be equally defined for a scalar wave field $\psi(\bm{r})$. 

As manifested in Eq.~\eqref{current} and~\eqref{poynting_vector} both of these quantities can be expressed in terms of the phase gradient given by 
\begin{equation}
    \boldsymbol\nabla \varphi({\bm r}) = \mathrm{Im}[\boldsymbol\nabla \log \psi({\bm r})] = \mathrm{Im}\left[\frac{\boldsymbol\nabla \psi({\bm r})}{\psi({\bm r})}\right].
    \label{phase_gradient}
\end{equation}
This gradient points in the direction where there is a maximum change in phase. Gradients are normal to contour surfaces, and hence the phase gradient is always normal to the wavefront \cite{Born_Wolf_optics}. The phase gradient $\boldsymbol\nabla\varphi({\bm r})$ is a vector field that can provide information about light propagation as $\boldsymbol\nabla \varphi({\bm r}) \varpropto {\bm k}({\bm r})$, with ${\bm k}({\bm r})$ the local wavevector, and the associated energy flow. 


We are mainly concerned with the energy flow on a plane perpendicular to the propagation axis of the beam, which dictates how a planar probe interacts with light. The associated phase gradient field reads
\begin{equation}
{\bm F}({\bm r}) \equiv \boldsymbol{\nabla} \varphi({\bm r}) - \frac{{\bm k}}{|{\bm k}|^2} {\bm k}\cdot \boldsymbol{\nabla} \varphi({\bm r}).
\label{projection}
\end{equation}
By applying the Helmholtz-Hodge Decomposition, discussed in Sec.~\ref{HHD}, we isolate the rotational energy flows or solenoidal phase gradients within the field \cite{Bahl12}, enabling us to distill the OAM content of the beam.

\section{Helmholtz-Hodge Decomposition}
\label{HHD}

The Helmholtz-Hodge Decomposition (HHD) is grounded in the Helmholtz theorem \cite{KU2020161}, which asserts that any field $\bm{F}$, defined in a region $\Omega$ on a bounded domain, can be segregated into an irrotational (curl-free) $\bm{f}_{\text{IR}}$ and a solenoidal (divergence-free) $\bm{f}_{\text{R}}$ component, determined by

\begin{equation}
    \bm{F}(\bm{r}) = \underbrace{\boldsymbol{\nabla} \Phi(\bm{r})}_{\equiv \bm{f}_{\text{IR}}(\bm{r})} + \underbrace{\boldsymbol{\nabla} \times \bm{A}(\bm{r})}_{\equiv \bm{f}_{\text{R}}(\bm{r})}\,,
\end{equation}
where $\Phi(\bm{r})$ and $\bm{A}(\bm{r})$ are the scalar and vector potentials {of $\bm{F}(\bm{r})$, respectively, not to be confused with the corresponding scalar and vector potentials of the electric and magnetic field. These quantities} can be obtained from Poisson's equations:

\begin{align}
    \Phi(\bm{r}) &= \frac{1}{4\pi} \int_V \frac{\boldsymbol\nabla \cdot \bm{F}(\bm{r}')}{|\bm{r} - \bm{r}'|} \, dV', \\
    \bm{A}(\bm{r}) &= \frac{1}{4\pi} \int_V \frac{\boldsymbol\nabla \times \bm{F}(\bm{r}')}{|\bm{r} - \bm{r}'|} \, dV'.
\end{align}

These potentials $\Phi(\bm{r})$ and $\bm{A}(\bm{r})$ allow the field $\bm{F}$ to be segregated into the curl-free and divergence-free components within a volume $V$. The boundary conditions that are imposed in this decomposition ensure a normal boundary flow on the curl-free component and a tangential flow on the divergence-free component. Considering $\hat{n}$ as the outward normal to the boundary $\Omega$, this implies that for a unique decomposition:
\begin{itemize}[leftmargin=1em]
    \setlength\itemsep{0.1em}
    \item The irrotational component is normal to the boundary $d\Omega$ of $\Omega$, i.e., $\bm{f}_{\text{IR}} \times \hat{\bm n} = 0$, and
    \item The solenoidal component is parallel to the boundary $d\Omega$ of $\Omega$, i.e., $\bm{f}_{\text{R}} \cdot \hat{\bm n} = 0$.
\end{itemize}

\section{Main Quantities: the definition of the R-Index and the solenoidal current}
\label{sec:R-index}

The R-Index represents the fraction of the rotational orbital angular energy density in a structured optical field. The magnitude of the rotational part of the decomposed vector field, $|{\bm f}_R({\bm r})|$, quantifies the total solenoidal currents or the azimuthal currents present in the field. In other words, the associated Poynting vector component ${\bm P}_{R}({\bm k}) = |{\bm f}_R({\bm r})| I({\bm r})/k$ determines the total circulating power within a cross section of area $A$ of the beam at a plane $z$.   
Mathematically, the R-index is defined as:

\begin{equation}
\begin{split}
    \text{R}(z) = \frac{\iint |{\bm f}_{\rm R}({\bm r})| I({\bm r}) \, dA}{\iint \left( |\boldsymbol{\bm f}_R({\bm r})| + |\boldsymbol{\bm f}_{IR}({\bm r})| \right) I({\bm r}) \, dA},
\end{split}
\label{R-index}
\end{equation}

Thus, it quantifies the fraction of the average solenoidal power over the average total power transport within the $z=$ const. plane. The R-Index serves as a valuable metric for determining the available intrinsic orbital energy flow in structured fields, offering insight into their suitability for various OAM-based applications. The total orbital flow density in a field is responsible for the beam divergence, self-diffraction, and transverse energy circulation. The R-index is therefore associated with the extracted azimuthal flow component that leads to the OAM of the beam. In other words, the R-index directly correlates to the number of photons that possess OAM and are available for interaction for a planar probe.

Notice, that the R-index is not a measure of the strength of the solenoidal current since it only relies on its fraction relative to the irrotational component. This means that the comparison of two beams in terms of their R-index does not lead to a conclusion on which beams has a larger amplitude of solenoidal currents. To resolve this in our analysis we additionally calculate the normalized solenoidal current strength reading 
\begin{equation}
    \bar{J}_s(z) = \frac{\iint |{\bm f}_{\rm R}({\bm r})|I({\bm r})~\mathrm{d}A}{\iint I({\bm r})~\mathrm{d}A},    
    \label{definition_Ps}
\end{equation}
which directly probes the amplitude of the solenoidal currents in the beam. Notice that this quantity is not dimensionless but it scales inversely proportional to the characteristic length scale of the spatial structure of the light field. For beams, $\bar{J}_s \propto w_0^{-1}$ implying that tighter beams involve stronger currents, due to the increased involved phase gradients, see Eq.~\eqref{current}.

\section{Numerical approach}
\label{sec:numerics}

The HHD of the phase gradient vector field was carried out by employing Fast Fourier Transform (FFT) techniques. The field ${\bm F}({\bm r})$ was first sampled on a uniform Cartesian grid by employing Eq.~\eqref{phase_gradient} and Eq.~\eqref{projection} for its extraction from $\psi({\bm r})$ and its on-plane projection respectively. The grid was chosen such as it corresponds to a FFT Discrete Variable Representation (DVR) \cite{LightCarrington2000} enabling the efficient calculation of derivatives and integrals. Since FFT introduces periodic boundary conditions the value of $\psi({\bm r})$ at the edges of our numerical grid were masked by Tukey windows to avoid artifacts. Notice that the mask of the Tukey windows addresses only regions of $I < 10^{-6} I_0$, which are subsequently dropped from our analysis.

The HHD was performed by numerically solving the Poisson equation
\begin{equation}
    \nabla^2 \Phi({\bm r}) = \boldsymbol{\nabla} \cdot {\bm F}({\bm r}),
\end{equation}
resulting to ${\bm f}_{\rm IR}({\bm r}) = \boldsymbol{\nabla} \Phi({\bm r})$ and ${\bm f}_{\rm R}({\bm r}) = {\bm F}({\bm r}) - \boldsymbol{\nabla} \Phi({\bm r})$. The right hand side $\boldsymbol{\nabla} \cdot {\bm F}({\bm r})$ can exhibit steep behavior close to the vortices, we resort to a FFT-based finite difference solver for the Poisson equation \cite{SchumannSweet1988} in order to avoid issues with spectral ringing. Within this approach the solution in Fourier space reads
\begin{equation}
\begin{split}
    \mathcal{F}[\Phi](k_x,k_y) = &\frac{1}{2} \left(\frac{\cos \frac{2 \pi k_x}{N_x}-1}{\Delta x^2} + \frac{\cos \frac{2 \pi k_y}{N_y}-1}{\Delta y^2}\right)^{-1}\\
    &\times \mathcal{F}[\boldsymbol{\nabla}\cdot{\bm F}](k_x,k_y), 
\end{split}
\end{equation}
where $k_{\mu}$, $N_{\mu}$, $\Delta \mu$, with $\mu =\{ x, y\}$ are the grid positions, number of grid points, grid spacing for the $\mu$ direction respectively and $\mathcal{F}$ represents the FFT. The boundary condition ${\bm f}_{\rm IR} \times \hat{\bm n} = {\bm 0}$ is enforced geometrically by the mirror symmetry of the source $\boldsymbol{\nabla} \cdot{\bm F}({\bm r})$ for $x \to -x$ and $y \to -y$ that we impose by appropriately choosing the orientation of the beam $\psi({\bm r})$ and the periodic boundary conditions naturally stemming from the FFT.

All calculations were performed in MATLAB. 
The FFT DVR employed consists of $4096$ points for each direction, while it extents between $|x|,|y| < 4 \sqrt{o} w_0$, where $o \geq 1$ is the maximum mode order of the Laguerre--Gaussian or Hermite--Gaussian components. The Tuckey windows affect the outer $20 \%$ of the grid.
The above choices provide sufficient resolution to capture both the fine structure of the field and allow for the smooth decay towards the boundaries.

\section{Results: Evaluation of R-Index across various field profiles}
\label{sec:results}

We conducted extensive numerical calculations to compute the R-index for diverse SSOF profiles, demonstrating its universal applicability in quantifying the OAM content. This quantity, defined as the intensity-weighted fraction of solenoidal optical currents, Eq.~\eqref{R-index}, serves as a direct metric for OAM accessibility in photons. To quantify the solenoidal field flow we also evaluate the normalized solenoidal current strength, $\bar{J}_s$, see Eq.~\eqref{definition_Ps}.

Throughout our analysis we use scaled units with respect to $w_0$ and $z_R$. Since we consider fixed $z/z_R$, the precise value of $\lambda$ for fixed $w_0$ does not affect our calculations except for a shift of the phase $\sim e^{-i k z}$, that affects neither $\bar{J}_s$ or the R-index.

Our key findings are summarized below:

%

        
         

\begin{table*}[t]
    \centering
    \renewcommand{\arraystretch}{1.5}
    \setlength{\tabcolsep}{0.01\linewidth} 
    
    \begin{tabular}{|
    >{\centering\arraybackslash}p{0.15\linewidth} | %
    >{\centering\arraybackslash}p{0.25\linewidth} | %
    >{\centering\arraybackslash}p{0.25\linewidth} | %
    >{\centering\arraybackslash}p{0.25\linewidth} |} %
    \hline
        \makecell{\textbf{SSOF profile}} & 
        \makecell{\textbf{Pure LG beam} \\
        ($l=1$ and $z = 0.1\,z_R$)} &
        \makegapedcells\makecell{\textbf{Gaussian beam} \\
        \textbf{with imprinted LG phase} \\
        ($l=1$ and $z = 0.1\,z_R$)} &
        \makecell{\textbf{Honeycomb optical lattice}} \\
        \hline
        \textbf{$\bar{J}_s$ (units of $w_0^{-1}$)} & $1.247$ & $1.856$ & $27.53\,(0.0964\,\lambda^{-1})$ \\
        \textbf{R-index} & $0.87$ & $0.11$ & $0.70$ \\
        \makecell{\textbf{Intensity profile}\\ \textbf{of SSOF}} &
        \makecell{\includegraphics[width=0.99\linewidth]{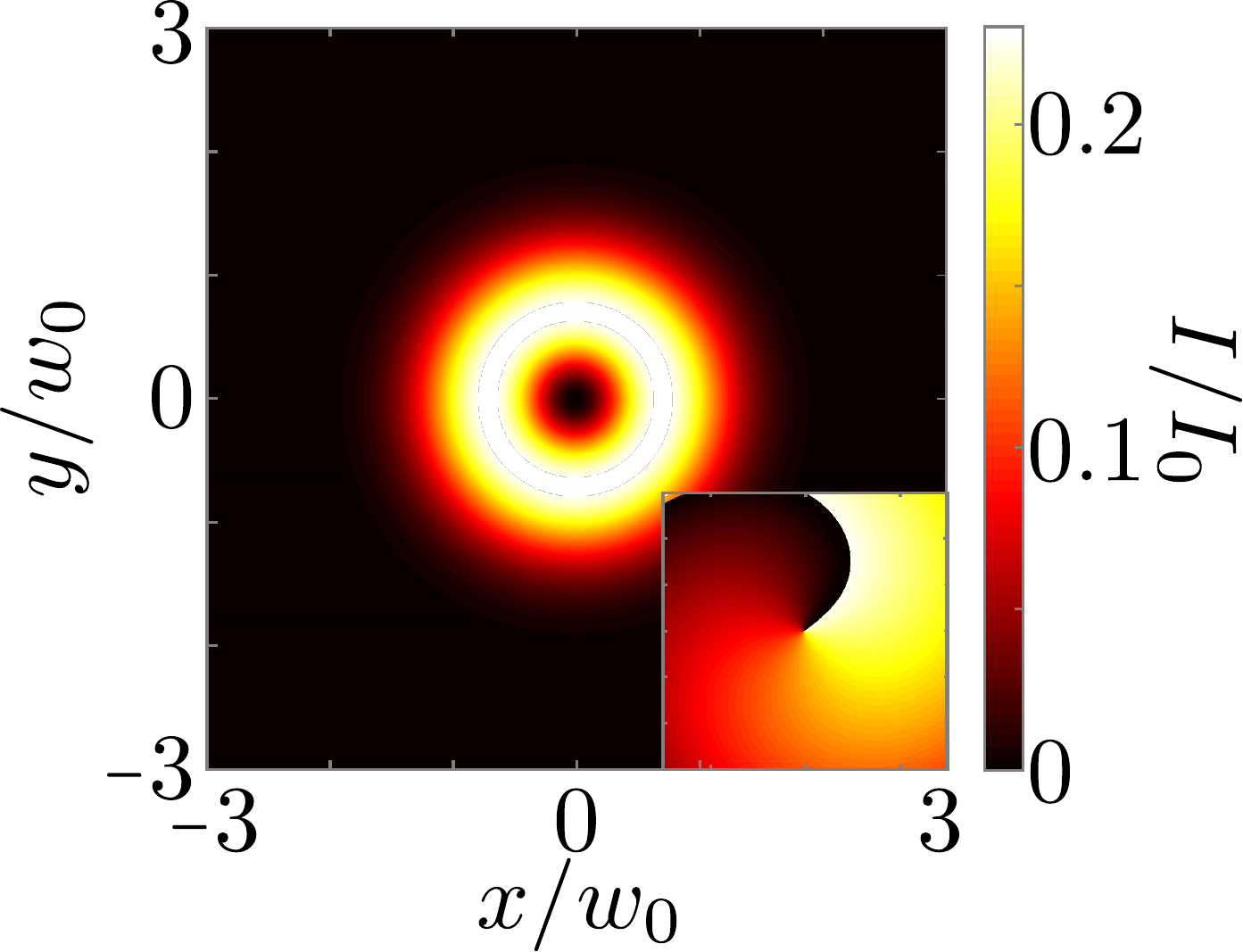}} & 
        \makecell{\includegraphics[width=0.99\linewidth]{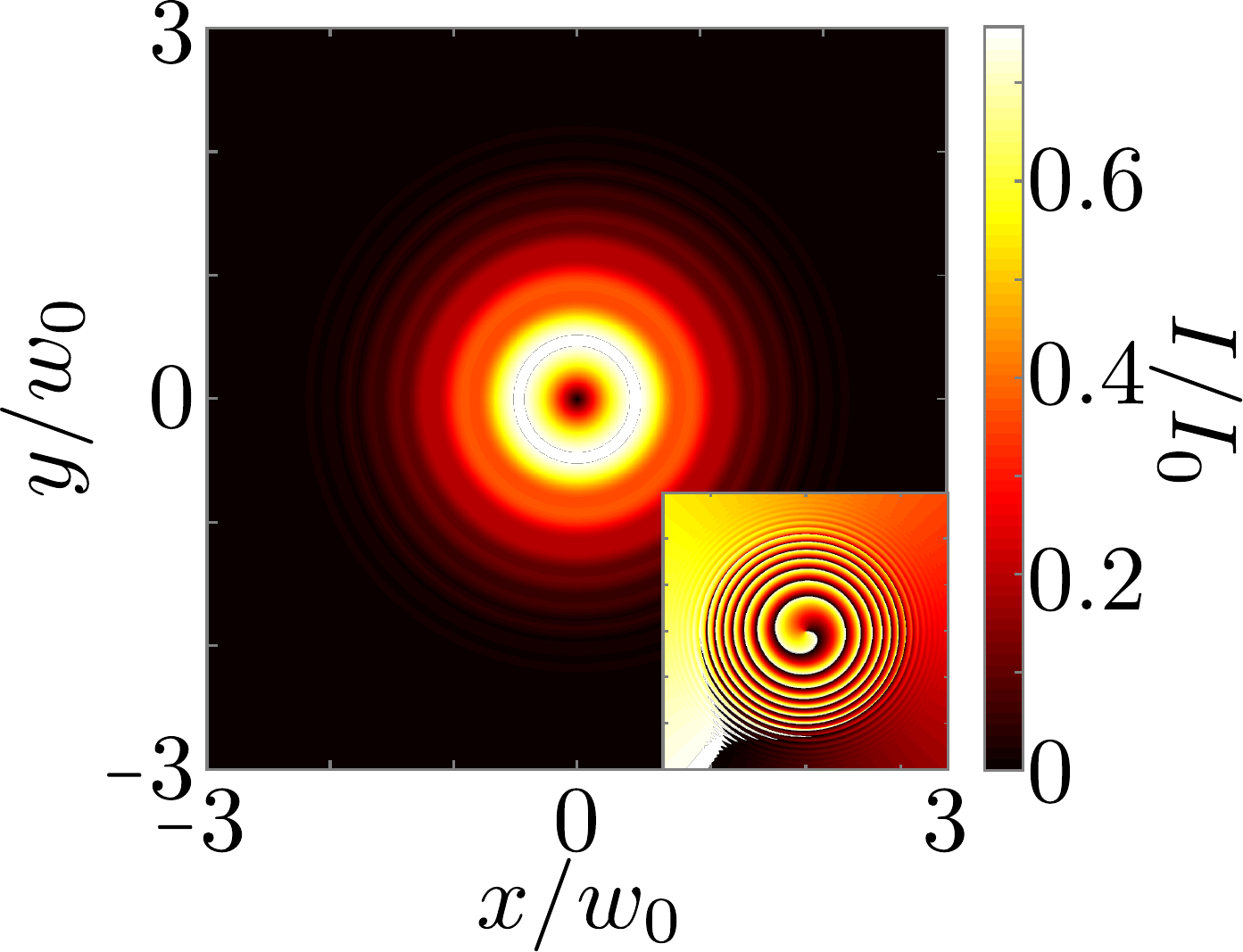}} & 
        \makecell{\includegraphics[width=0.99\linewidth]{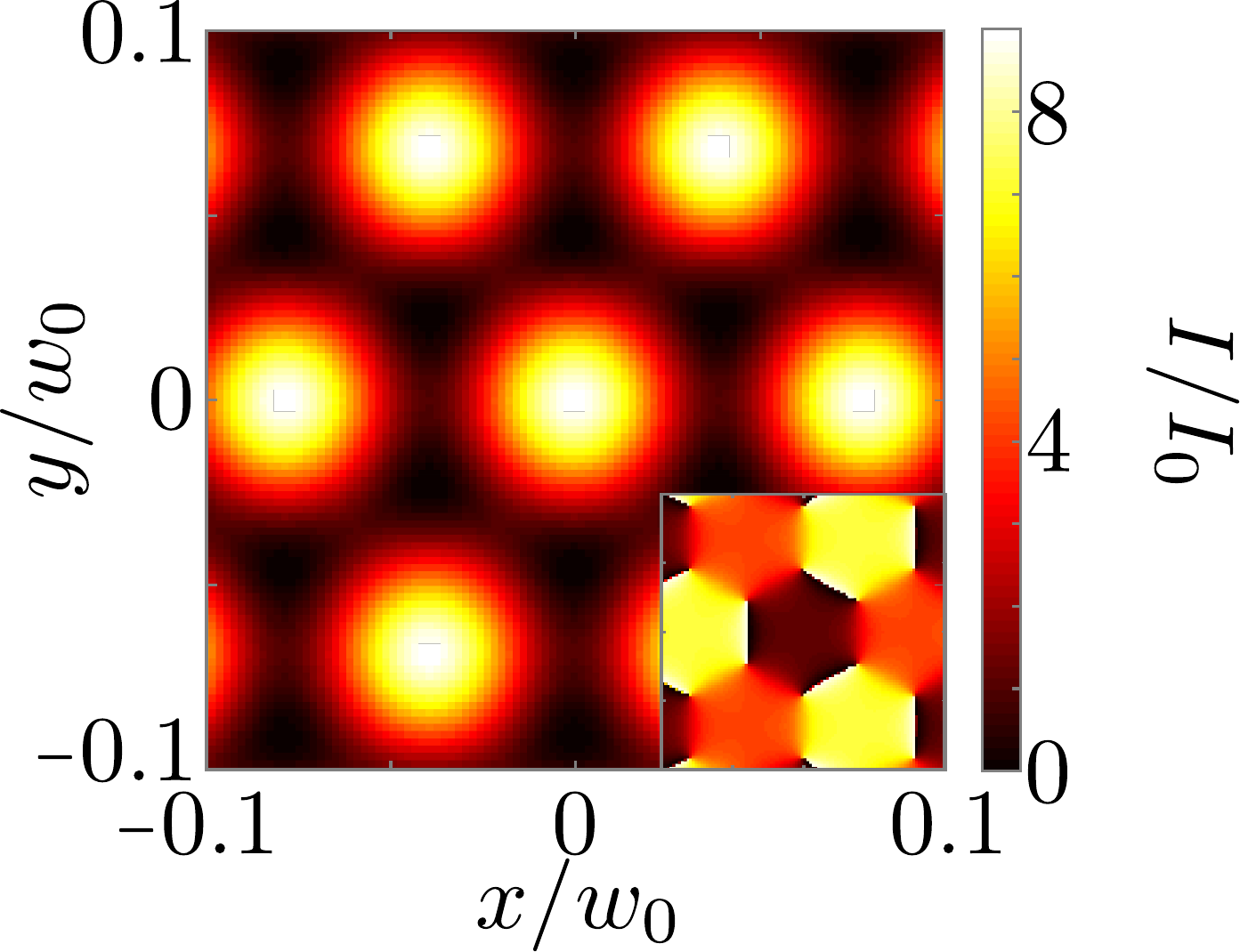}}\\
    \hline
    \end{tabular} 
    \caption{Numerically obtained OAM content characteristics for three example SSOF profiles. The OAM is characterized in terms of the R-index and $\bar{J}_s$ and the corresponding intensity pattern (the phase appears in the inset) is additionally provided. For the comparison of the beam and lattice $\bar{J}_s$ and phase $\lambda = 0.0035 w_0$ is considered.} 
    \label{tab_SSOF_Comparison}
\end{table*}

\emph{Pure Laguerre-Gaussian (LG) beams}-- The profile of these beams reads
\begin{adjustwidth}{1.65em}{0pt}
\begin{equation}
\begin{split}
\psi_{p,l}^{\rm LG}({\bm r}) &= E_0 \frac{w_0}{w(z)} 
        \left( \frac{\sqrt{2} r}{w(z)} \right)^{\lvert l\rvert}
        L_p^{|l|}\left( \frac{2r^2}{w(z)^2} \right) \\
        &\quad \times \exp\left( -\frac{r^2}{w(z)^2} \right) \exp\left( -i \frac{kr^2}{2R(z)} \right) \\
    &\quad \times 
    \exp(-ikz) \exp(i\xi(z)) \exp(il\phi),
\label{LG}
\end{split}
\end{equation}
\end{adjustwidth}
where $E_0$ is the field amplitude, $w_0$ the beam waist, $w(z) = w_0 \sqrt{1 + (z/z_R)^2}$, $z_R = \pi w_0^2/\lambda$ is the Rayleigh range, $R(z) = z [1 + (z_R/z)^2]$ is the radius of curvature, and $\xi(z) = (|l| + p + 1) \tan^{-1}(z/z_R)$ is the Gouy phase.

\begin{figure}[t]
    \centering
    \includegraphics[width=0.95\columnwidth]{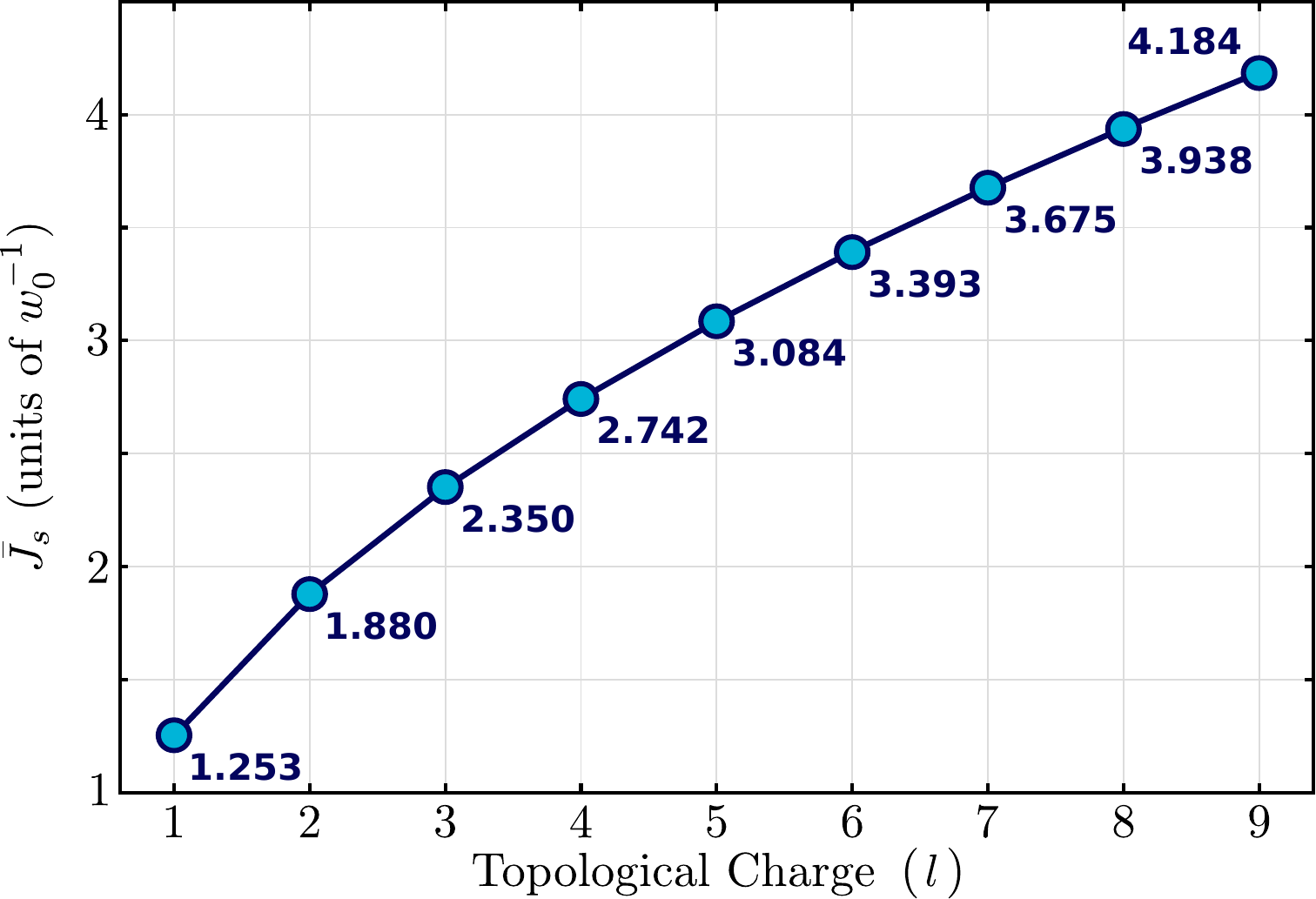}
    \caption{The normalized solenoidal current strength, $\bar{J}_s$, as a function of the topological charge for LG beams ($p =0$ and $l=1$ to $9$) for $z = 0$.}
    \label{fig:solenoidal_currents}
\end{figure}

Exactly at the focus $z = 0$, these beams have a purely azimuthal ${\bm F}({\bm r})$, leading to a purely solenoidal current since $\boldsymbol{\nabla} \cdot {\bm F}({\bm r}) = 0$, and hence an R-index of $1$. The strength of this solenoidal current increases with the topological charge $\bar{J}_s \sim \sqrt{2 l}$ as revealed in Fig.~\ref{fig:solenoidal_currents}. However, as the beam propagates away from its focus its divergence leads to additional radial currents that cause the reduction of the R-index. Details on the $z/z_R$ dependence of the R-index for Laguerre-Gaussian beams are discussed in Sec.~\ref{sec:analytics}.
For comparison to other field profiles, here, we consider a beam propagated to $z = 0.1~z_R$.
Table \ref{tab_SSOF_Comparison} reveals that such a beam with $p=0$ and $l = 1$ achieves an R-index of 0.87 thus retaining near-ideal OAM purity (87\% rotational energy flow).

The fraction of solenoidal orbital optical current density values were computed for LG beams with varying topological charges, revealing the increasing relationship (see Table~\ref{tab:vortex_metrics} and Fig.~\ref{fig:solenoidal_currents}). 

\vspace{1ex}
\begin{table}[h!]
    \centering
    \renewcommand{\arraystretch}{1.1}
    \setlength{\tabcolsep}{5pt}
    \footnotesize
    \begin{tabular}{c c c}
        \toprule
        \textbf{Topological} & \textbf{$\bar{J}_s$ units of $w_0^{-1}$} & \textbf{R-index} \\
        \textbf{Charge $l$}      & & \\
        \midrule
        1 & 1.247  & 0.869 \\
        2 & 1.871  & 0.889 \\
        3 & 2.339  & 0.896 \\
        4 & 2.728  & 0.899 \\
        5 & 3.069  & 0.901 \\
        6 & 3.376  & 0.902 \\
        7 & 3.657  & 0.903 \\
        8 & 3.918  & 0.904 \\
        9 & 4.163  & 0.905 \\
        \bottomrule
    \end{tabular}
    \caption{Magnitude of extracted solenoidal phase gradient $\bar{J}_s$ and R-index for LG beams (topological charges 1--9) at $z = 0.1 z_R$ from focus.}
    \label{tab:vortex_metrics}
\end{table}

\emph{Phase--imprinted Gaussian beam (Hypergeometric Gaussian beams)}--
A usual way that vortex beams of expreimentally implementing vortex beams by imprinting the azimuthal phase of a Laguerre--Gaussian beam on a Gaussian mode. This is because Guassian modes can be generated with very high fidelity \cite{YangWang2021, YangDan2025} and phase imprinting can be achieved with readily available methods such as spiral phase plates or spatial light modulators (SLM) \cite{SuedaMiyaji2004}. To consider such a phase imprinting process we the ideal case for both the focussed Gaussian beam with the additional spiral phase corresponding to $l = 1$
\vspace{-1.2ex}
\begin{adjustwidth}{1.65em}{0pt}
\begin{equation}
\begin{split}
\psi^{\rm IM}(r,\phi,z = 0) &= E_0 \exp\left(-\frac{r^2}{w_0^2}\right) e^{i \phi}.
\end{split}
\label{G}
\end{equation}
\end{adjustwidth}
\vspace{-1.0ex}
The solution of the paraxial equation with the above field at the focus $z = 0$ is known to be an example of a Hypergeometric Guassian mode (HyGG) and in particular a modified Bessel-Gauss mode~\cite{KarimiGianluigi2007}. The field of this beam reads
\begin{equation}
\begin{split}
\psi^{\rm IM}(\tilde{r}, \phi, \tilde{z}) =& E_0\frac{\sqrt{\pi}}{2} \frac{\tilde{r}}{\sqrt{\tilde{z}}(\tilde{z}+i)^{3/2}} \exp\left(- \frac{i \tilde{r}^2}{\tilde{z}} \right)\\
&\bigg[I_0\left(\frac{\tilde{r}^2}{2 \tilde{z}(\tilde{z}+i)} \right) - I_1\left(\frac{\tilde{r}^2}{2 \tilde{z}(\tilde{z}+i)} \right)\bigg] \\ 
    &\exp\left(-\frac{\tilde{r}^2}{2 \tilde{z}(\tilde{z}+i)} + i \phi - i k z\right),
\end{split}
\end{equation}
where we have introduced the scaled coordinates $\tilde{z} = z/z_R$ and $\tilde{r} = r/w_0$.
A distinctive property of this beam is the factor $e^{-i \tilde{r}^2/\tilde{z}}$ corresponding to a steep phase gradient in the direction away from the focus. This gradient correspond to a radial current ${\bm j}({\bm r})$ that forces the intensity away from $r = 0$ where the optical vortex is imprinted.

Table~\ref{tab_SSOF_Comparison} manifests that while the solenoidal current $\bar{J}_s$ is similar in amplitude to the case of the vortex beam, the R-index is very small taking a value of $0.11$. This behavior occurs exactly because of the large radial current described above which contributes to the irrotational part of the phase gradient, ${\bm f}_{\rm IR}({\bm r})$.

The above shows that the R-index is a sensitive probe of the purity of the generated OAM mode. By quantifying SPP-generated beam purity, the R-index provides a tool for optical vortex generation protocols. Its measurement by state-of-the-art phase measurement approaches such as phase-shifting holography~\cite{AndersenAlperin2019}, provides a detection pathway for optimizing fabrication tolerances of optical elements and alignment protocols. This is particularly crucial for mid-infrared applications where spatial light modulators are unavailable necessitating the development of novel optical vortex creation techniques. 


\begin{table*}[t]
    \centering


    \renewcommand{\arraystretch}{1.5}
    \setlength{\tabcolsep}{0.01\linewidth} 

    \begin{tabular}{|
    >{\centering\arraybackslash}p{0.2\linewidth}| %
    >{\centering\arraybackslash}p{0.22\linewidth}| %
    >{\centering\arraybackslash}p{0.22\linewidth}| %
    >{\centering\arraybackslash}p{0.22\linewidth}| %
    }
        \hline
        \textbf{SSOF Profile} &
        \textbf{HG$_{\bm{2,0}}$ + HG$_{\bm{0,2}}$} &
        \textbf{HG$_{\bm{2,0}}$ + HG$_{\bm{0,2}}$} &
        \textbf{HG$_{\bm{1,3}}$ + HG$_{\bm{3,1}}$} \\
        \hline
        \textbf{Phase Difference} &
        $\pi/2$ &
        $\pi/4$ &
        $\pi/2$ \\
        \textbf{$\bar{J}_s$ units of $w_0^{-1}$} &
        1.410 &
        1.383 &
        1.653 \\
        \textbf{R-index} &
        0.74 &
        0.66 &
        0.83 \\
        \makecell{\textbf{Intensity Profile}} &
        \makecell{\includegraphics[width=0.59\linewidth]{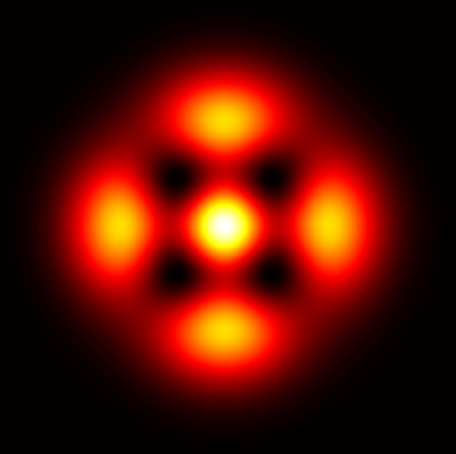}} &
        \makecell{\includegraphics[width=0.59\linewidth]{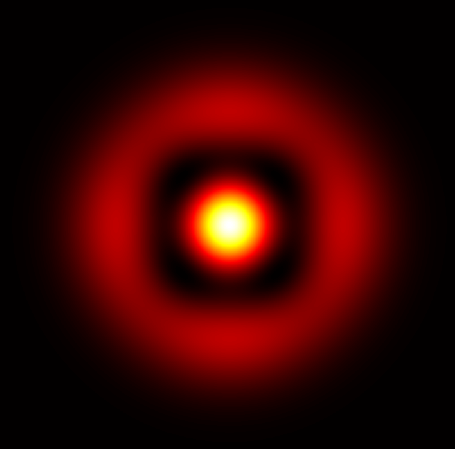}} &
        \makecell{\includegraphics[width=0.59\linewidth]{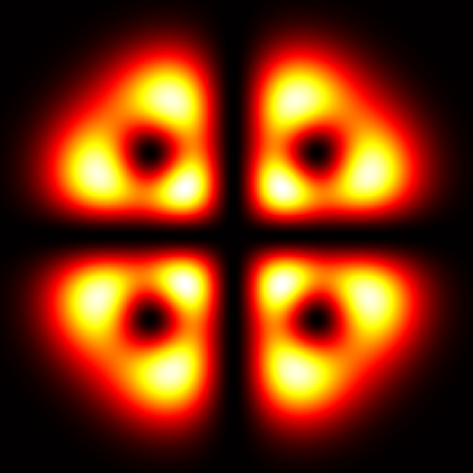}} \\
        \hline
    \end{tabular}
    \caption{Comparison of different SSOFs generated by superpositions of Hermite-Gaussian modes. The OAM is characterized in terms of the R-index and $\bar{J}_s$ and the corresponding intensity patern (and phase in the case of phase imprinted Gaussian profile) is additionally provided. In all cases the field at the focus $z = 0$ is considered.}
    \label{tab:tab_ssofs_mt}
\end{table*}

\emph{Honeycomb optical lattice}-- An interference of three plane waves produces an optical vortex lattice, the field profile of which reads 
\begin{adjustwidth}{1.65em}{0pt}
\begin{equation}
\begin{split}
\psi^{\rm L}({\bm r}) = E_0 \sum_{j=1}^3 \exp \Big[
    -i k \big( &\sin\theta \cos\phi_j\, x \\
    + &\sin\theta \sin\phi_j\, y \\ 
    + &\cos \theta\, z\big)
\Big],
\end{split}
\label{array}
\end{equation}
\end{adjustwidth}
where $k = 2\pi/\lambda$, $\theta = 1.7^\circ$ is the angle of the plane-waves with respect to $z$-axis, and $\phi_j$ are the corresponding azimuthal angles ($\phi_1 = 120^\circ$, $\phi_2 = 240^\circ$, $\phi_3 = 360^\circ$).

This lattices exhibits an R-index of $0.70$, when measured in the $z = 0$ plane, see Table~\ref{tab_SSOF_Comparison}.
This structure yields a large local solenoidal current as indicated by $\bar{J}_s$, but fragmentation of dark cores leads to low precentage of OAM content. Notice that in the case of the lattice $\bar{J}_s$ is only dependent on $\lambda$. Here the value of $\bar{J}_s$ in terms of $w_0^{-1} = 0.0035 \lambda^{-1}$ is provided for comparison of the lattice and beam currents.

\emph{Hermite-Gaussian (HG) superpositions}--
Hermite–Gaussian beams of order $m, n$, $\mathrm{HG}_{m,n}$, are described by the well-known expression:

\begin{adjustwidth}{1.65em}{0pt}
\begin{equation}
\begin{split}
\psi^{\mathrm{HG}}_{m,n}({\bm r}) = \, 
& E_0 \frac{w_0}{w(z)} H_m\left(\frac{\sqrt{2}x}{w(z)}\right)
  H_n\left(\frac{\sqrt{2}y}{w(z)}\right) \\
& \times  \exp\left(-\frac{x^2 + y^2}{w(z)^2}\right)\\
&\times \exp\left(-i k \frac{x^2 + y^2}{2R(z)}\right) \\
& \times 
  \exp\big(-i \xi(z)\big)
  \exp(-i k z).
\end{split}
\label{HGmode}
\end{equation}
\end{adjustwidth}

where $H_m$ and $H_n$ are Hermite polynomials of orders $m$ and $n$, and $w(z)$, $R(z)$, and $\xi(z) = (m + n +1) \tan^{-1}(z/z_R)$ denote the beam radius, radius of curvature, and Gouy phase, respectively. The expressions for $w(z)$ and $R(z)$ are identical to the case of Laguerre--Gaussian beams, see Eq.~\eqref{LG}.

Complex modes exhibiting optical vortices are formed by superpositions of such HG modes~\cite{Ohtomo:08,Chu:12} with varied phase differences introduced between them.

HG superposed profiles showed reduced R-index values, underscoring mode interference as a critical purity-limiting factor. It is evident that the phase difference between superposed modes significantly influences the R-index, as anticipated. For example, Table~\ref{tab:tab_ssofs_mt} shows that a superposition of $\mathrm{HG}_{2,0}$ and $\mathrm{HG}_{0,2}$ modes with a phase difference of $\pi/2$ yields an R-index of $0.74$, which decreases to $0.66$ when the phase difference is modified to $\pi/4$. In comparison, a superposition of $\mathrm{HG}_{1,3}$ and $\mathrm{HG}_{3,1}$ modes with a phase difference of $\pi/2$ achieves an R-index of $0.83$, indicating that nearly $83\%$ of the photons in this configuration are available for OAM-based experimental applications.

Table~\ref{tab:tab_ssofs} in Appendix~\ref{app:HGSSOFs} presents R-index values for a more diverse collection of structurally singular optical fields (SSOFs) formed through Hermite-Gaussian mode superpositions.

Our analysis demonstrates that Laguerre--Gaussian beams are optimal for efficient OAM transfer to planar targets. However, realizing high-purity LG beams necessitates the use of high-fidelity beam generation, which can be challenging, particularly in the mid-infrared (Mid-IR) spectral range. Given the absence of spatial light modulators (SLMs) in the Mid-IR and the current limitations in continuous SPP fabrication, it may be advantageous in some cases to employ superpositions of HG modes or structured alternatives like vortex arrays which become viable substitutes, as they achieve comparable intrinsic OAM densities despite variations in spatial profiles and mode compositions. This flexibility enables adaptable OAM system design. The R-index thus serves as a fundamental metric for field optimization, enabling performance equivalent substitutions.

\vspace{0.5cm}
\noindent
\begin{minipage}{\columnwidth}
\centering
{\textbf{\textit{Critical Insights}}}
\end{minipage}

\begin{itemize}
\item \textbf{Universal purity quantification:} The R-index enables direct comparison of OAM content across arbitrary fields, from single vortices to complex superpositions, addressing a longstanding gap in structured light characterization.

\item \textbf{Phase-imprinting performance validation:} By quantifying the beam purity of beams generated by phase manipulation, the R-index provides a tool for optimizing experimental vortex creation methodology. This is particularly crucial for mid-infrared applications where spatial light modulators are unavailable.

\item \textbf{Design optimization:} The metric identifies LG modes as theoretical optima while revealing HG superpositions as viable alternatives when LG generation is impractical, guiding experimental design choices.
\end{itemize}

\section{Analytical Calculation of the R-index for LG Beams}
\label{sec:analytics}


To analytically corroborate our numerical findings in this section we analytically evaluate the R-index for all Laguerre--Gaussian beams, Eq.~\eqref{LG}, with $p = 0$ and arbitrary $l$. 

The intensity profile of these beams is
\begin{equation}
I({\bm r}) = I_0 \left( \frac{w_0}{w(z)} \right)^2 \left( \frac{\sqrt{2} r}{w(z)} \right)^{2|l|} 
\exp\left( -\frac{2 r^2}{w(z)^2} \right),
\end{equation}
with $I_0 = \frac{c \epsilon_0}{2} |E_0|^2$,
and the phase profile reads
\begin{equation}
\varphi = -\frac{kr^2}{2R(z)} - kz + \xi(z) + l\phi.
\end{equation}
The phase gradient can be analyzed in simple terms when expressed in cylindrical coordinates
\begin{equation}
\boldsymbol\nabla \varphi = \frac{\partial \varphi}{\partial r} \hat{\bm r} + \frac{1}{r} \frac{\partial \varphi}{\partial \phi} \hat{\bm \phi} + \frac{\partial \varphi}{\partial z} \hat{\bm z}.
\end{equation}
By taking the curl and divergence of each component in the above expression, it is easy to verify that only the azimuthal component is of solenoidal character, ${\bm f}_{\rm R}({\bm r}) = \frac{l}{r}\hat{\bm \phi}$. The radial component is irrotational, ${\bm f}_{\rm IR}({\bm r}) = -\frac{kr}{R(z)}\hat{\bm r}$, while the $z$ component does not contribute to ${\bm F}({\bm r})={\bm f}_{\rm IR}({\bm r}) + {\bm f}_{\rm R}({\bm r})$ due to its projection on the perpendicular plane, see Eq.~\eqref{projection}.


\begin{table*}[t]
    \centering
    \renewcommand{\arraystretch}{1.2} 
    \setlength{\tabcolsep}{5pt} 
    \small

    \begin{tabular}{|
    >{\centering\arraybackslash}p{3cm}| %
    >{\centering\arraybackslash}p{2cm}| %
    >{\centering\arraybackslash}p{3cm}| %
    >{\centering\arraybackslash}p{2cm}| %
    >{\centering\arraybackslash}p{2cm}| %
    }
        \hline
        \textbf{SSOF Profile} & \textbf{Phase Difference} & \textbf{$\bar{J}_s$ units of $w_0^{-1}$} & \textbf{R-index} & \textbf{Intensity Profile} \\ 
        \hline
        HG$_{3,0}$ + HG$_{0,3}$ & $\pi/2$ & 1.549 & 0.71 & \adjustbox{valign=c}{\includegraphics[width=13mm]{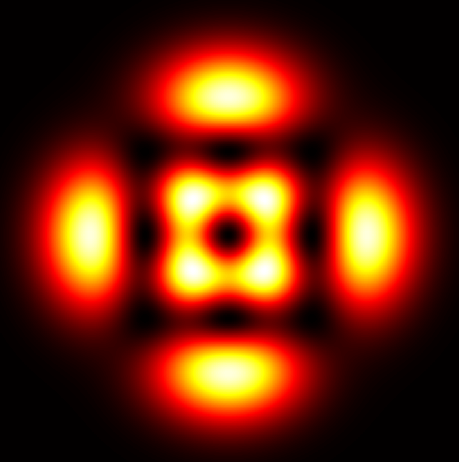}}  \vspace{2pt} \\
        HG$_{3,0}$ + HG$_{0,3}$ & $\pi/4$ & 1.556 & 0.62 & \adjustbox{valign=c}{\includegraphics[width=13mm]{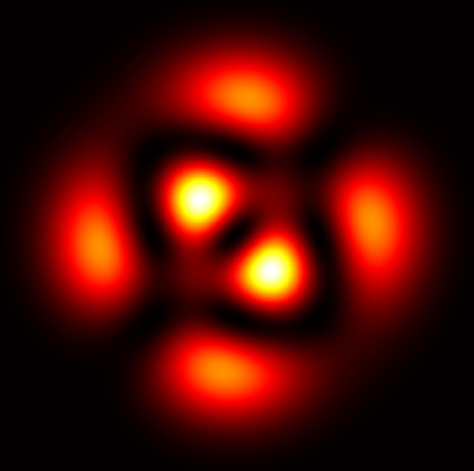}}  \vspace{2pt} \\
        HG$_{1,2}$ + HG$_{2,1}$ & $\pi/2$ & 2.227 & 0.73 & \adjustbox{valign=c}{\includegraphics[width=13mm]{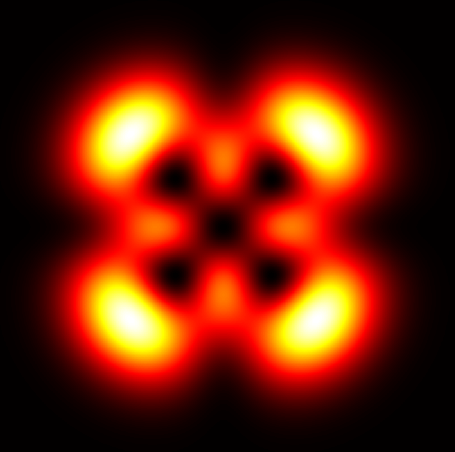}}  \vspace{2pt} \\
        HG$_{1,2}$ + HG$_{2,1}$ & $\pi/4$ & 1.636 & 0.67 & \adjustbox{valign=c}{\includegraphics[width=13mm]{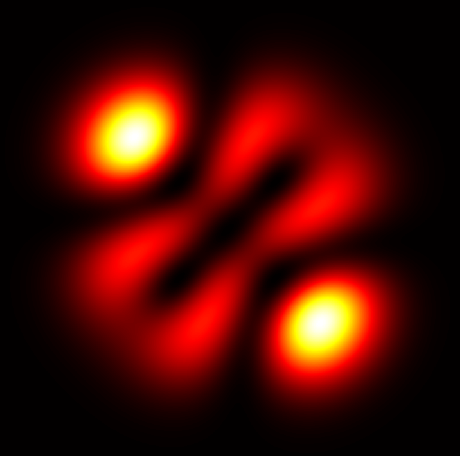}}  \vspace{2pt} \\
        HG$_{1,3}$ + HG$_{3,1}$ & $\pi/4$ & 1.643 & 0.67 & \adjustbox{valign=c}{\includegraphics[width=13mm]{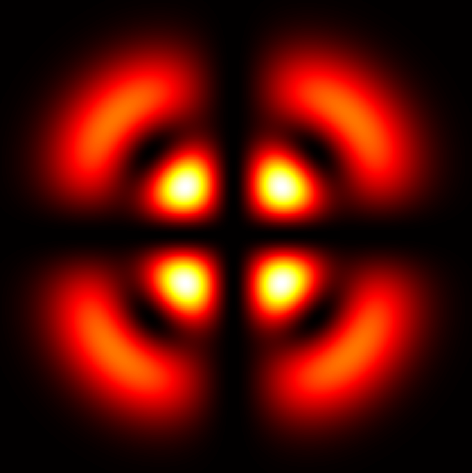}}  \vspace{2pt} \\
        HG$_{1,3}$ + HG$_{3,1}$ & $\pi/3$ & 1.643 & 0.75 & \adjustbox{valign=c}{\includegraphics[width=13mm]{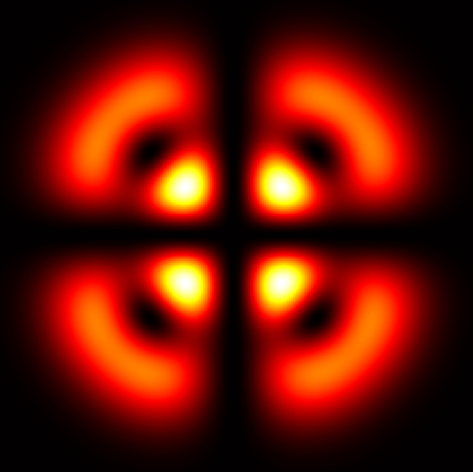}}  \vspace{2pt} \\
        HG$_{2,4}$ + HG$_{4,2}$ & $\pi/2$ & 2.665 & 0.69 & \adjustbox{valign=c}{\includegraphics[width=13mm]{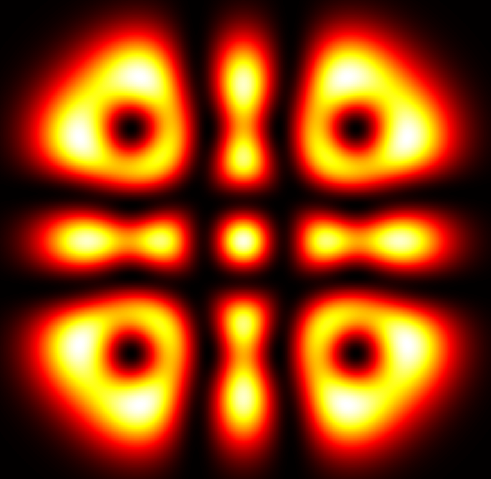}}  \vspace{2pt} \\
        HG$_{2,4}$ + HG$_{4,2}$ & $\pi/4$ & 2.636 & 0.61 & \adjustbox{valign=c}{\includegraphics[width=13mm]{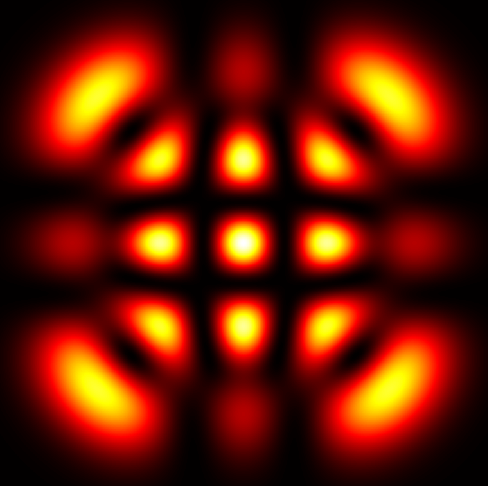}}  \vspace{2pt} \\
        HG$_{2,4}$ + HG$_{4,2}$ & $\pi/3$ & 2.647 & 0.65 & \adjustbox{valign=c}{\includegraphics[width=13mm]{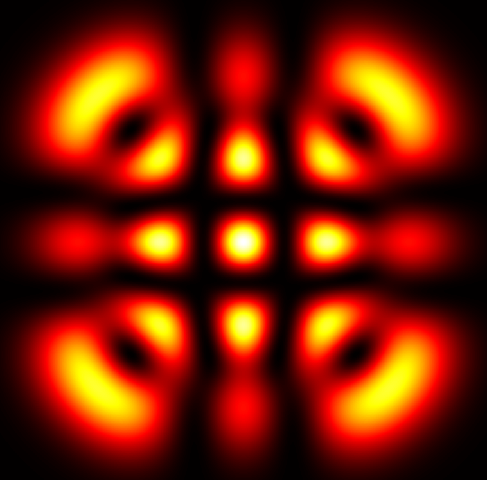}} \vspace{2pt} \\
        HG$_{3,2}$ + HG$_{2,3}$ & $\pi/2$ & 2.840 & 0.74 & \adjustbox{valign=c}{\includegraphics[width=13mm]{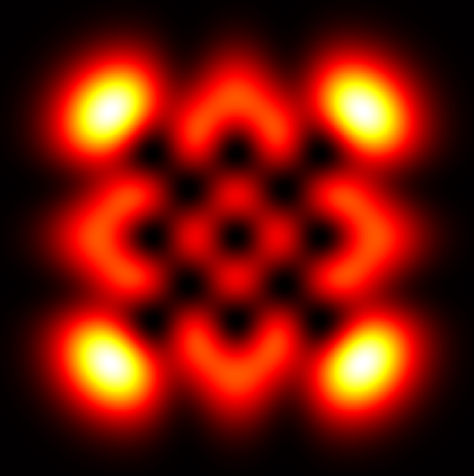}} \vspace{2pt} \\
        HG$_{3,2}$ + HG$_{2,3}$ & $\pi/4$ & 2.822 & 0.64 & \adjustbox{valign=c}{\includegraphics[width=13mm]{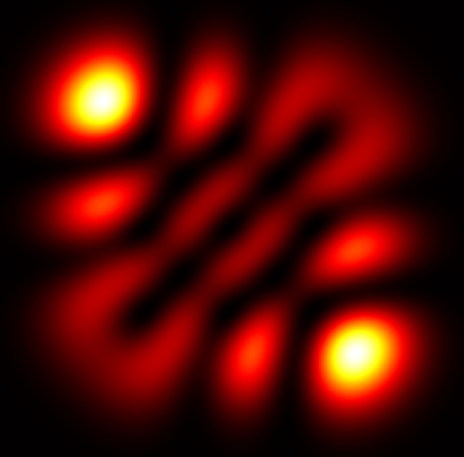}} \vspace{2pt} \\
        HG$_{3,4}$ + HG$_{4,3}$ & $\pi/2$ & 3.320 & 0.74 & \adjustbox{valign=c}{\includegraphics[width=13mm]{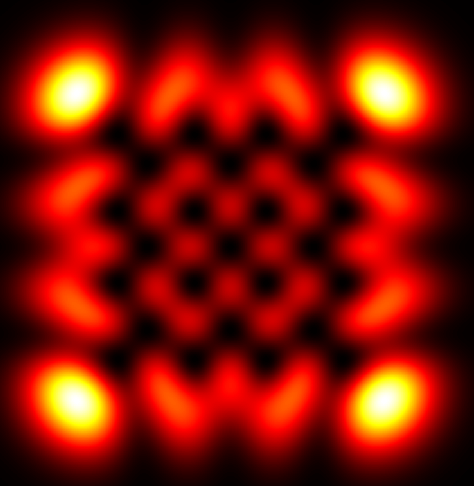}} \vspace{2pt} \\
        \hline
    \end{tabular}
    \caption{Same as Table~\ref{tab:tab_ssofs_mt} but for an extended selection of Hermite-Gaussian superpositions.}
    \vspace{-0.72cm}
    \label{tab:tab_ssofs}
\end{table*}

The R-index of Eq.~\eqref{R-index} can then be calculated analytically by employing the integral~\cite{Gradshteyn}
\begin{equation}
\int_0^{\infty} r^{2 k} e^{-\frac{2 r^2}{w^2}} \, {\rm d}r =
    \dfrac{(2k)!}{2^{3k +1} k!} \sqrt{\frac{\pi}{2}} w^{2k +1},
\label{integral}
\end{equation}
%
The numerator straightforwardly evaluates to
\begin{equation}
\begin{split}
\iint |{\bm f}_{\rm R}({\bm r})| I({\bm r}) \, \mathrm{d}A =& \sqrt{\frac{\pi^3}{2}} \frac{|l| (2|l|)!}{2^{2|l|} |l|!}\\
&\times \frac{I_0 w_0}{\sqrt{1 + \left( z/z_R \right)^2}}.
\end{split}
\label{numerator}
\end{equation}
The integral for $|{\bm f}_{\rm IR}({\bm r})|$ can also be straightforwardly calculated
\begin{equation}
\begin{split}
    \iint|{\bm f}_{\rm IR}({\bm r})|I({\bm r})~\mathrm{d}A =& \sqrt{\frac{\pi^3}{2}} \frac{(2|l|+2)!}{2^{2|l|+2} (|l|+1)!} \\
    &\times \frac{I_0 w_0}{\sqrt{1 + \left( z/z_R \right)^2}} \frac{z}{z_R}.
\end{split}
\label{denominator}
\end{equation}
%
Therefore, by using Eq.~\eqref{numerator} and Eq.~\eqref{denominator} the R-index yields
\begin{equation}
{\rm R}(z) = \left(1 + \frac{2 |l| +1}{2 |l|} \frac{z}{z_R}\right)^{-1}
\end{equation}

This expression reveals that at the focus, $z = 0$, of any Laguerre-Gaussian beam the propagation of light is purely solenoidal and thus a planar probe would identify purely OAM bearing current. However, as the probed plane moves away from the focus the OAM content decreases as the beam diverges. The decrease is more pronounced for beams of smaller $l$, e.g. $l =1$ yields $\mathrm{R}(z) \approx 1 - \frac{3}{2} \frac{z}{z_R}$, while for larger $l \gg 1$, the decrease is smoother tending towards $\mathrm{R}(z) \approx 1 - \frac{z}{z_R}$. This behavior independently verifies the numerical results of Table~\ref{tab:vortex_metrics} and can be used for the determination of the $l$ quantum numbers of beams of low $l$.

\section{Conclusion}
\label{sec:conclusion}
In this work, we introduced the R-index, a robust metric to quantify the intrinsic OAM content of spatially structured optical fields. It establishes a unified framework for quantifying intrinsic OAM content. By directly measuring the relative intensity-weighted solenoidal energy, the R-index provides a quantitative assessment of mode purity and the fraction of photons available for OAM-mediated interactions, even in complex fields with multiple vortices.

Unlike traditional approaches based on vortex counting or modal decomposition, the R-index decouples beam purity from topological charge and provides a quantitative measure of mode purity. It offers a robust tool for enabling empirical optimization of OAM sources. Our calculations demonstrate its applicability across a range of beam types, including Laguerre-Gaussian, Gaussian, spiral phase plate-generated, and Hermite-Gaussian beams, making it a universal tool for structured light analysis. To our knowledge, this is the first demonstration of a universal OAM quantification method based on a single index, enabling direct comparison of disparate field structures, including LG beams and complex multi-vortex superpositions, on an equal footing. This facilitates the identification of the most OAM-efficient configuration, regardless of field complexity. Future directions include extending the R-index framework to vectorial fields, experimental validation through interferometry or modal decomposition, and exploring its behavior under beam propagation or turbulence.

\section*{Acknowledgments}
This research was funded in whole or in part by the Austrian Science Fund (FWF) [10.55776/F1004]. For open access purposes, the author has applied a CC BY public copyright license to any author accepted manuscript version arising from this submission.

\section*{Data Availability Statement}
Data underlying the results presented in this paper are not publicly available at this time but may be obtained from the authors upon reasonable request.




\appendix

\section{Solenoidal phase gradient magnitudes and R-index values for various SSOFs}
\label{app:HGSSOFs}

This appendix includes the Table~\ref{tab:tab_ssofs}
providing an extensive array of examples of SSOFs with their associated $\bar{J}_s$ and R-index values.

%
%

\bibliography{bibliography}

\end{document}